\documentstyle[11pt]{article}    
\begin{document}  
 
\newcommand{\bra}{\langle}     
\newcommand{\ket}{\rangle}       
\def\tr{{\rm tr}\,} 
\def\href#1#2{#2}

\begin{titlepage} 
 
\begin{center} 
\hfill HUTP-01/A046\\ 
\hfill UOSTP-01106\phantom{a\,} \\ 
\hfill hep-th/0110039 
\vspace{2cm} 
 
{\Large\bf Supersymmetric Brane-Antibrane Configurations}

\vspace{1.5cm} 
{\large 
Dongsu Bak$^a$ and  Andreas Karch$^b$ } 
 
\vspace{.7cm} 
 
$^a${\it 
Physics Department, University of Seoul, Seoul 130-743, Korea 
\\ [.4cm]} 
$^b${\it 
Jefferson Physical Laboratory, 
Harvard University, 
Cambridge, MA 02138, USA\\ 
[.4cm]} 
({\tt dsbak@mach.uos.ac.kr, karch@fresnel.harvard.edu}) 
 
\end{center} 
\vspace{1.5cm} 
 
 
We find a class of flat supersymmetric  
brane-antibrane configurations.  
They follow from ordinary brane-antibrane systems by 
turning on a specific worldvolume background  
electric field, which corresponds to dissolved fundamental strings.  
We have clarified in detail how they arise and 
identified their constituent charges as well as the 
corresponding supergravity solutions. 
Adopting the matrix theory description, 
we  construct  the worldvolume gauge theories and  
prove the absence of any tachyonic degrees. We also study  
supersymmetric  solitons of the worldvolume theories.     
\vspace{3.5cm} 
\begin{center} 
\today 
\end{center} 
\end{titlepage} 
 
\section{Introduction} 
 
Recently a  new class of supersymmetric brane configurations 
has been discovered in string theory, the so called supertubes. 
They were originally constructed as 1/4 BPS solutions to the  
Born-Infeld action of a D2 brane with worldvolume gauge fields turned 
on \cite{mateos}. Many supertubes and their worldvolume theory have been  
investigated  using 
matrix theory in \cite{lee} and a corresponding supergravity 
solution was constructed in \cite{emparan}. The supertube 
is a cylindrical D2 brane which is prevented from collapse 
by angular momentum. The configuration has 
zero D2 brane charge but carries a D2 dipole moment, as expected 
from a tube like configuration. The reason it can be BPS is 
that the worldvolume fields  turned on, which carry 
the angular momentum, induce D0 and F1 brane charge, which ultimately 
become the charges  appearing in the corresponding supersymmetry algebra. 
More recently, generalizations including junctions of  tubes 
have been studied in Ref.\cite{swkim}  
 
In this paper we want to explore a much simpler system that exhibits 
a similar physics: a parallel configuration of D2 and anti-D2 brane 
in flat space can be made 1/4 BPS by turning on  worldvolume B and E 
fields. This configuration can be obtained by deforming the circular 
tube into an elliptic tube and by taking the limit where the ellipse 
degenerates into two parallel lines. The separation 
between the branes is a free parameter. Both branes have the same E-field 
and opposite B-field. The fact that this deformation into an ellipse 
still preserves supersymmetry can easiest be seen in the matrix theory description 
of the tube, where the BPS equations for the matrices $X$, $Y$ and $Z$ encoding 
the positions of the constituent D0 branes can be brought into the form \cite{lee}:  
\begin{equation} 
[X,Y]=0, \; \; \; [X,[X,Z]] + [Y,[Y,Z]] =0. 
\label{bpseq} 
\end{equation} 
This is solved by setting $[X,Z] =il Y$ and $[Z,Y] =il X$. 
For the circular tube the same constant $l$ has been chosen in both equations, 
but in order to solve the BPS equations one only needs that $[X,Z] \sim Y$ and 
$[Z,Y] \sim X$, so 
 different constants can be chosen. This 
corresponds to the deformation into the ellipse.  
 
At first it may sound counterintuitive that a brane and an antibrane can 
be BPS together. The way this works is that in the presence of both E and 
B-field, the supersymmetry preserved by a D2 brane is the same as the supersymmetry 
of an anti D2-brane, as long as the B-fields come with opposite signs. This 
will be shown in Section 2 by studying the kappa symmetry on the D2 
worldvolume. Once we established that D2 and anti-D2 preserve the same SUSY 
it is clear that one can actually have an arbitrary number of D2 and anti-D2 
branes at arbitrary positions in the transverse space with or without 
net D2 brane charge and still describe a stable 1/4 BPS configuration. 
Even the magnitude of the B-field can vary from brane to brane as long 
as we keep the sign choice correlated with brane or antibrane. In 
section 3 we write down the supergravity solution for these supersymmetric brane 
configurations and analyze T-dual setups. 
 
Section 4 is devoted to a study of the D2 anti-D2 system using matrix 
theory. We once more demonstrate that the configuration preserves 1/4 of 
the supersymmetry. Analyzing small fluctuations we 
show explicitly that by turning on the electric  
field on the worldvolume 
the tachyon in the D2 anti-D2 system disappears.  
The worldvolume theory is non-commutative SYM. 
We again construct configurations corresponding 
to arbitrary superpositions of branes. We also exhibit 
solutions corresponding to branes at angles. 
In Section 5 we study the worldvolume theory 
using the open string metric. Possible decoupling limits are discussed. 
Section 6 is devoted to a study of worldvolume solitons.

\section{Worldvolume action and kappa-symmetry} 
 
For a D2 brane to be supersymmetric, one has to find Killing spinors 
of the background geometry, which we take to be just flat space, that satisfy 
\begin{equation} \label{gamma} 
\Gamma \epsilon = \pm \epsilon 
\end{equation} 
where $\Gamma$ is the matrix appearing in the worldvolume kappa symmetry 
action that depends on the embedding, the type of brane 
and the worldvolume fields that are turned on. The upper 
or lower sign refers to brane and antibrane respectively. So it seems 
obvious that the spinors that satisfy the equation with the plus sign can't 
satisfy the equation with the minus sign simultaneously. This 
conclusion can be avoided, if we turn on different worldvolume 
fields on D2 and anti-D2 respectively, so that the left side picks 
up an extra sign as well. What 
we like to show is that in the presence of an worldvolume $E$-field 
1/4 of the supersymmetries are preserved as long as one turns on $B$-fields 
of opposite sign on D2 and anti-D2 branes. As we will see, the magnitude 
of B actually doesn't matter. 
 
In the case of a D2 brane along $txz$ with $F_{xz} =B$ and $F_{zt}=E$ 
turned on, the matrix $\Gamma$ becomes \cite{bergshoeff}: 
\begin{equation} \label{gamdtwo} 
\Gamma= \frac{\sqrt{\det(g)}}{\sqrt{\det(g+ 2 \pi \alpha' F)}} (\gamma_{txz} 
+ E\; \gamma_x \gamma_{11} + 
 B\; \gamma_t \gamma_{11}). 
\end{equation} 
In what follows we will take the background metric to be 10d flat space. 
For analyzing possible scaling limits we will 
later restore the constants $g_{tt}$, $g_{xx}$ and 
$g_{zz}$,  
but for now we just work with a metric $g_{\mu \nu} =\eta_{\mu \nu}$. 
Similarly, we will frequently set $ 2 \pi \alpha'=1$. 
The $\gamma$ matrices are the induced worldvolume Dirac 
matrices. Since we are dealing with a flat brane in flat space, 
they are just equal to their embedding space counterparts. 
 
For $E=B=0$, Eq.~(\ref{gamdtwo}) reduces to 
\begin{equation} \label{dtwo} 
\gamma_{txz} \epsilon = \pm \epsilon 
\end{equation} 
which is the usual condition for a D2 along $txz$. D2 and anti-D2 
preserve opposite supersymmetries. 
 
What we do corresponds to solving (\ref{gamdtwo}) by setting 
\begin{eqnarray} 
\label{one} 
(\gamma_{txz} + E \gamma_x \gamma_{11}) \epsilon &=&0            \\ 
\label{two} 
(B \gamma_{t} \gamma_{11} \mp \frac{\sqrt{g+ 
 F}}{\sqrt{g}}) \epsilon &=&0    
\end{eqnarray} 
Notice that the sign that differs for D2 and anti D2 only makes it into 
(\ref{two}), not into (\ref{one}). As we will see momentarily, 
imposing those two conditions simultaneously preserves $1/4$ 
of the supersymmetries. 
 
Eq. (\ref{one}) is solved by setting $E=-1$\footnote{One can choose  
either signatures of $E$. But for the later comparison,  
we here choose the negative one.}  
and imposing 
\begin{equation} 
\label{solveone} 
\gamma_{tz} \epsilon = - \gamma_{11} \epsilon. 
\end{equation} 
In the absence of a magnetic field, $|E|=1$ would be 
the critical value of the electric field, where the fundamental 
string becomes tensionless, $\sqrt{g+F}$ vanishes. Since we 
turned on a $B$ field in addition, $|E|=1$ isn't critical in the 
sense that the tension of strings goes to zero. 
For a single  
D2 brane, we could go to a frame where 
only an electric field 
or only a magnetic field is 
turned on depending on whether the Lorenz invariant quantity
$E^2-B^2$ 
is positive or negative.
As the magnitude of $B$ is arbitrary, we can 
choose   $E^2-B^2$ to have either sign.
 But even 
if we choose it to be positive with a nonvanishing $B$ field
in the original frame, in the new frame with $E$ field only,  
$|E_{new}| <1$. 
Since we will soon be dealing with 
many branes, all with the same $E$, but with different $B$s, we will 
stay in the original frame where both are non-zero. What does happen 
at $|E|=1$ is that 
\begin{equation} \label{critical} 
\sqrt{g+F} = \sqrt{1+B^2-E^2} = \sqrt{B^2} 
\end{equation} 
which we will still loosely refer to as a ``critical'' electric field in what  
follows. 
 
Eq.~(\ref{two}) is solved by 
\begin{equation} 
\label{solvetwo} 
 \gamma_t \gamma_{11} \epsilon = \pm \frac{\sqrt{B^2}}{B} \epsilon 
\end{equation} 
where again the two different signs refer to D2 brane and 
anti-D2 brane respectively. Now it is obvious that as long as we choose 
the sign of B to be positive for a D2 brane and negative for an anti-D2 
brane, Eq.~(\ref{solvetwo}) in both cases reduces just to 
\begin{equation} 
\label{dzero} 
\gamma_t \gamma_{11} \epsilon = \epsilon 
\end{equation} 
This is just the supersymmetry preserved by a D0 brane. Similarly 
(\ref{solveone}) is the supersymmetry preserved by a fundamental 
string along $t$ and $z$. We see that the supersymmetry of this 
configuration is only sensitive to the constituents, that is the 
lower brane charges induced by the background fields. For D0's and 
F1's we know that the conditions (\ref{solveone}) and (\ref{dzero}) 
are consistent with each other and preserve $1/4$ of 
the supersymmetries (the D0-F1 system is dual to D3-D5 or 
D1-F1). 
 
But now since both D2 and anti-D2 with 
the appropriate sign of the $B$-field preserve the  
same supersymmetries (those associated with F1 
and D0) it is clear that there is no force and we can have configurations 
with an arbitrary number of D2 and an arbitrary number of anti-D2's 
at arbitrary positions in the  
transverse directions. Note that the magnitude of the B-field 
is allowed to differ from brane to brane, as long as the sign 
choice is correlated with the brane being D2 or anti-D2.

\section{Supergravity solutions} 
 
The supergravity solution for a single D2 brane with the E and 
B field of the kind we discussed turned on was constructed 
in \cite{emparan} as a limit of the supertube metric, zooming in to the region 
very close to the tube where the tube looks planar. The metric in this limit 
reads 
\begin{equation} 
\label{metric} 
ds^2 = - U^{-1} V^{-1/2} (dt - k dx)^2 + U^{-1} V^{1/2} dz^2 + 
        V^{1/2} dx^2 + V^{1/2} ( dr^2 + r^2 d \Omega_6^2) 
\end{equation} 
for $N$ coincident D2 branes  
along $t$, $x$ and $z$, and the harmonic functions 
$U$, $V$ and $k$ 
being given by the usual expressions for  
smeared F1, D0 and D2 brane charge along $t$, $x$ and $z$ 
localized at the origin $r=0$ in the transverse seven space: 
\begin{eqnarray} 
\label{harmonic} 
U&=& 1 + \frac{12 \pi^3 g_s^2 l_s^6 n_1}{r^5}  \\ \nonumber 
V&=&  1+ \frac{24 \pi^4 g_s l_s^7 n_0}{r^5}  \\ \nonumber 
k&=& \frac{6 \pi^2 g_s l_s^5 N}{r^5}  
\end{eqnarray} 
where $n_0$ and $n_1$ are the D0 and F1 charge densities in string units 
respectively. 
The other nonzero SUGRA fields are given by  
\begin{eqnarray} 
B_{(2)} &=& - U^{-1} (dt - k dx) \wedge dz + dt \wedge dz \\ \nonumber 
C_{(1)} &=& - V^{-1} (dt- k dx) + dt \\ \nonumber 
C_{(3)} &=& - U^{-1} k dt \wedge dz \wedge dx \\ \nonumber 
e^{\phi} &=& U^{-1/2} V^{3/4} 
\end{eqnarray} 
giving rise to a 4-form field strength 
\begin{equation} 
G_{(4)} = d C_{(3)} - dB_{(2)} \wedge C_{(1)} =   
U^{-1} V^{-1}  dt \wedge dz  
\wedge dx \wedge dk 
\end{equation} 
appropriate for $N$ D2 branes at $r=0$. 
Since this configuration is BPS, as shown in \cite{emparan}, one can 
construct many-centered brane solutions with branes at 
$\vec{y}_a$ by just superposing 
harmonic functions with $\frac{1}{r^5}$ replaced by  
$\frac{1}{|\vec{y} - \vec{y}_a|^5}$. While only positive 
contributions can be added to $U$ and $V$, $k$ can receive both 
positive and negative contributions corresponding to a D2 or an anti-D2 
respectively 
located at $\vec{y}_a$.  
 
As in the case of the supertube we can find a bound on the number 
of D2 brane charge in the system. Basically every D2 brane and anti-D2 
brane we introduce comes with a fixed amount of D0 and F1 charge. 
If for a given amount of D0 and F1 brane charge we want to maximize 
the D2 brane charge, we should have no antibranes and no extra unbounded
 D0's or F1's. 
Every other configuration has less D2 brane charge. This way one obtains 
an upper bound on $N$ in terms of $n_0$ and $n_1$. 
As in the case of the tube in the gravity solution this bound 
can be reproduced by studying closed time like curves \cite{emparan}. 
If $g_{xx} <0$ in addition to $g_{tt} < 0$ we can find 
a continuous path $x^\mu(s)$ for $s \in [0,1]$ such that 
\begin{equation} 
x^\mu (s=0)=x^\mu (s=1) 
\end{equation} 
and 
\begin{equation} 
g_{\mu\nu}\, {dx^\mu\over ds} {dx^\nu\over ds} < 0 
\end{equation} 
for all $s \in [0,1]$. 
For this not to happen, we have to require 
\begin{equation} 
- U^{-1} V^{-1/2} k^2 + V^{1/2} \geq 0 
\end{equation} 
which translates into 
\begin{equation} 
\label{bound} 
N^2 \leq n_0 n_1 (2 \pi)^3 g_s l_s^3. 
\end{equation} 
 
From this solution it is straightforward to T-dualize to higher dimensional 
branes, that is Dp and anti-Dp branes which are 1/4 BPS because 
they carry (smeared) D(p-2) and F1 charge. 
To do this one has to implement the following changes: 
\begin{itemize} 
\item the harmonic functions now only depend on the radial 
coordinate in the $9-p$ dimensional transverse space, their falloff 
being given by $\frac{1}{r^{7-p}}$. 
\item the formerly transverse directions that become  
worldvolume directions upon T-duality come with a $V^{-1/2}$  
instead of the $V^{1/2}$. Otherwise the metric is the same as above. 
\item the R-R forms $C_{(3)}$ and $C_{(1)}$ get replaced with 
$C_{(p)}$ and $C_{(p\!-\!2)}$ by wedging the old solution 
with the new worldvolume directions. 
\item the dilaton changes under T-duality, and the new dependence 
is  
\begin{equation} 
e^{\phi} = U^{-1/2} V^{\frac{3-(p-2)}{4}} 
\end{equation} 
\end{itemize} 
 
It is a little more difficult to understand what happens if we T-dualize 
along one of the worldvolume directions of the two branes. 
In the case of the supertube this T-duality leads to a 
superhelix configuration, see \cite{cho,lunin}. 
One thing we can do is to go back to the worldvolume 
action and choose a gauge, where 
the worldvolume gauge field is 
\begin{equation} 
\label{gaugepot} 
A_z = B x - E t 
\end{equation} 
and all other components zero. Under T-duality 
a worldvolume gauge field of the D2 translates into the position in the 
transverse direction of the T-dual D1. We see that our D2 brane 
T-dualizes into a D1 string whose $z$ coordinate is given 
by (\ref{gaugepot}), that is it is moving with a speed $E$ in the z 
directions and rotated by $B$. Since D2 and anti-D2 branes have opposite signs 
of $B$ they are rotated away from each other.

\section{Matrix theory}                                                         
So far we have discussed the supersymmetric  
 D2 and anti-D2 systems from the view points of the  
supergravity or the Born-Infeld descriptions.  
 In this section, we shall exploit their  
properties employing the matrix theory description. 
As mentioned earlier, the supersymmetric  
D2 and anti-D2 system may be obtained from the  
noncommutative supersymmetric  
tubes\cite{lee,swkim} as a limiting case of the elliptic  
deformation. We will first describe  
the details 
of such deformation within the matrix theory  
description. A constituent D2 brane obtained  
this way preserves only a quarter of the total 32  
supersymmetries due to the presence of the worldvolume  
gauge field. We shall provide a detailed study of 
the worldvolume gauge theory that is  
noncommutative.  
Unlike the ordinary ${\rm D2}$-$\overline{\rm D2}$ which is  
tachyonic, there should not be any tachyonic degrees 
in our case. This will be proved using the noncommutative  
worldvolume gauge theory. 
We then provide more  
general supersymmetric 
brane configurations like branes at angles.

Let us begin with the matrix    
model  Lagrangian\cite{banks1}  
\begin{equation}     
L={1\over 2 R} \tr \left( \sum_I (D_0 X_I)^2    
+{1\over (2\pi \alpha')^2} 
\sum_{I<J} [X_I,X_J]^2+ {\rm fermionic\ part}   
\right)    
\label{lag}     
\end{equation}     
where $I,J=1,2,\cdots 9$, $R=g_s l_s$ is the    
radius of the tenth    
spatial direction, and $\alpha'(\equiv l_s^2)$ is  
related to 
 the eleven dimensional   
Planck length by $l_{11}= 
(R \alpha')^{1\over 3}$. 
The scale $R$ (together with  $2\pi\alpha'$)  
will be frequently omitted below by setting them unity.   
One could introduce a target space metric $g_{\mu\nu}$ 
into the action, which may be a convenient way to get  
the decoupling limit related to the noncommutative  
field theory. However, for simplicity, we shall not go into 
this complication in this section. Related to the descriptions  
below, one thing we like to emphasize is that the above 
model is valid for any finite $R$ and $\alpha'$.   
One does not need any further decoupling limit for the validity 
of the description\cite{seiberg1}. 
  
For the supersymmetric tubes or ${\rm D2}$-$\overline{\rm D2}$ 
systems, we shall turn on the first three components, $X$, $Y$ 
and $Z$.    Using the Gauss law, the 
Hamiltonian can be written in 
a complete square form plus commutators terms as 
\begin{equation}    
H={1\over 2 } \tr \left( (D_0 X+ i[Z,X])^2+(D_0 Y + i[Z,Y])^2+(D_0 Z)^2  
+|[X,Y]|^2+ 2 C_J  
\right)\  \ge\   \tr C_J  
\label{hamiltonian}    
\end{equation}    
where the central charge    $\tr C_J$ is defined by 
\begin{equation}    
\tr C_J= i\,\tr\Bigl( [X, Z (D_0 X)]+[Y, Z (D_0 Y)]\Bigr)\,.   
\label{centralcharge}    
\end{equation} 
Note that the central charge here is a trace of commutator  
terms. Hence for any finite dimensional representations, 
the central charge vanishes. As will be shown later on, 
this central charge is related to the stretched strings in  
the z-direction\cite{banks}.      
The relevant BPS equations 
were identified in Refs.\cite{lee,swkim}; in  
the gauge $A_0={1\over 2\pi \alpha'}Z$, 
they are given by Eq.(\ref{bpseq})   
with all the components static, i.e.  
$\partial_0 X=\partial_0 Y=\partial_0 Z=0$.  
Notice that the supersymmetric tube solutions in Refs.\cite{lee,swkim}  
satisfy  the  algebra,  
 $ [Z,X]=il Y$, $[Y,Z]=i l X$ and   $[X,Y]=0$   
where $l$ is an arbitrary parameter related to  
the noncommutativity scale. 
The representations of the algebra describe  tubes 
extended in $z$-direction. In particular, $\rho^2\equiv X^2+Y^2$  
is a Casimir 
of the algebra and proportional to the identity for any irreducible 
representations; the solutions describe 
 circular shaped tubes. There seem many variations of the algebra  
that lead to the solutions of the BPS equations. 
One simple deformation of interest is given by  
\begin{eqnarray}    
  [Z,X]=ia Y, \ \ [Y,Z]=i b X\,, \ \   [X,Y]=0,   
\label{bpssolutions}    
\end{eqnarray}    
with $a$ and $b$ arbitrary. One can  easily show that 
the corresponding configuration describes an elliptic tube and  
the ellipse is described by a Casimir, $\rho^2={1\over a}X^2 + {1\over b}Y^2$. 
We then take the degenerate limit where the scale $a$ becomes large 
while keeping $b$ and $\rho$ fixed. The resulting configuration is describing 
two separated planar D2 branes. Since the tube carries no D2-brane 
charges, it is clear that the resulting two brane configuration  
corresponds to 
a brane-antibrane system.  
As will be shown shortly, the limit is in fact described by 
the reduced BPS equations 
\begin{eqnarray}    
[X,Y]=[Z,Y]=0\ \ \ \  [X,[X,Z]]=0\,.  
\label{bps}    
\end{eqnarray}   
Any nontrivial solutions of these equations  will be 
again  
 1/4 BPS. This one may see as follows. 
Note that the supersymmetric variation of the  fermionic   
coordinates $\psi$ in the matrix theory is     
\begin{eqnarray}     
  \delta \psi =  \left(D_0 X^I\, \gamma_{I} + {i\over 2}[X^I,X^J] 
\,\gamma_{IJ}\right)\epsilon' +\tilde\epsilon\,,   
\label{susy}     
\end{eqnarray}     
where $\epsilon'$ and $\tilde\epsilon$ are   
real spinors of 
16 components parameterizing total 32 supersymmetries. 
These are related to the 11 dimensional  
32 component real spinor $\epsilon$ 
by  
\begin{eqnarray}     
  \epsilon'=\gamma_{11}\Omega^{11}_+ \epsilon\,,\ \ \ 
  \tilde\epsilon=\Omega^{11}_- \epsilon
\label{susy11}     
\end{eqnarray}    
with the projection operators  
$\Omega^{11}_\pm={1\over 2}(1\pm \gamma_t\gamma_{11})$. 
Using the BPS equations in (\ref{bps}), the invariance condition becomes 
\begin{eqnarray}     
D_0 X (\gamma_{x} + \gamma_{xz})\epsilon'  
+\tilde\epsilon=0\,,   
\label{susy12}     
\end{eqnarray}     
For the nontrivial configurations, this implies that 
\begin{eqnarray}     
 \Omega_+\epsilon'=0\,,\ \ \  \tilde\epsilon=0\,,   
\label{susy13}     
\end{eqnarray}     
where $\Omega_\pm \equiv {1\over 2}(1\pm\gamma_z)$ are  
projection operators. 
These two conditions agree respectively with (\ref{solveone}) 
and (\ref{dzero}) 
of the kappa symmetry consideration. 
The kinematical supersymmetries parametrized by $\tilde\epsilon$  
are completely broken while only half of remaining dynamical  
supersymmetries are left unbroken.    
Hence in total the configuration preserves a quarter    
of the 32 supersymmetries of the matrix model and the unbroken 
supersymmetries are the same as those  of the  
supersymmetric tubes. 
 
Before delving into the case of the 
supersymmetric ${\rm D2}$-$\overline{\rm D2}$,  
we will first construct a supersymmetric D2 with electric flux  from the   
 BPS equations in (\ref{bps}) and study its charge and worldvolume dynamics.  
A D2 with electric flux is described by the Heisenberg algebra 
\begin{equation} 
[x,z]=i\theta,\ \ y=0 
\end{equation} 
where $\theta$ is the noncommutativity  
parameter. As usual one may find the representation of 
this algebra introducing  the annihilation and creation operators,  
 $c$ and $c^\dagger$, by 
\begin{equation} 
c={1\over \sqrt{2\theta}} (x+iz)\,,\ \ \ 
c^\dagger={1\over \sqrt{2\theta}} (x-iz)  
\end{equation} 
with $[c,c^\dagger]=1$. The minimal 
irreducible  
representation 
of the algebra will then be 
\begin{eqnarray} 
x+iz=\sqrt{2\theta} c= 
\sqrt{2\theta}\sum^\infty_{n=0}\sqrt{n+1} |\,n\,\ket\bra n+1|\,. 
\end{eqnarray} 
This background  
describes D0's distributed uniformly in the $x$-$z$ plane.  
The description respects the rotational symmetry around the 
origin. The operator  
$r^2\equiv x^2+z^2$ is diagonalized by the states $|n\ket$ 
with eigenvalues $\theta (n+1/2)$.  

For the charges, we will use the nonabelian Chern-Simons   
couplings of D-particles to the R-R gauge fields\cite{myers},  
\begin{equation}     
S_{\rm CS}=  
\mu_0 \int dt\, \tr \left( C^{(1)}_t  
+C^{(1)}_{I} D_t \phi^I+{i\lambda\over 2}   
C^{(3)}_{t\,IJ}[\phi^J,\phi^I]+{i\lambda^2\over 3}  
\phi^I\phi^J\phi^K F^{(4)}_{t\,IJK}+{\rm h.o.t.}\right)  
\,,    
\label{charges}     
\end{equation}      
where $\mu^{-1}_p=(2\pi)^p g_s\, l_s^{p+1}$, $\lambda=2\pi \alpha'$,  
$X^I=2\pi \alpha' \phi^I$ and $F^{(p+1)}$ is the field strength  
corresponding to the R-R p-form potential, $C^{(p)}$.   
The first term implies that the charges of D0 is counted by 
$\tr I$.  
The third term  
 implies there is now net  
$D2$-brane charge.  
The last term vanishes and 
there is no dipole moment that couples to  the R-R four  
form field strength, partly because the D2 is located at the origin 
($\vec{y}=0$) 
in the transverse space. 
 
Noting  
\begin{equation}     
S^{\rm D2}_{\rm CS}=  
{i\mu_0\lambda\over 2 } \int\, dt\, \tr  
[\phi^J,\phi^I] 
C^{(3)}_{t\, IJ}=  
{1\over(2\pi)^2  g_s \,\,l_s^3}   
\int dt dx dz   
\, C^{(3)}_{txz}  
\,,    
\end{equation}  
 we see that the D2 charge density is given by $\mu_2$ 
as expected.  
For the density of D0's on the D2 brane, we use the relation 
$\int dxdz= 2\pi \theta \tr I $ where $\tr I$ corresponds to  
the number of D0's as said above. Thus the number density 
$n_0\equiv \tr I/\int dxdz$ is   
\begin{equation}     
n_0={1\over 2\pi \theta}\,. 
\end{equation}         
The fundamental strings are stretched in the  
z directions producing a 
worldvolume electric field. To evaluate the corresponding  
number density $n_1\equiv 
N_s/\int dx$, we note that the central charge\cite{banks,swkim} 
is related to the number of strings $N_s$ by 
\begin{equation}     
\int dt\, \tr C_J= {1\over 2\pi \alpha'} N_s \int dt dz\,. 
\end{equation}    
Hence one finds that 
\begin{equation}     
n_1= {\theta\over (2\pi)^2 g_s l^3_s}\,. 
\end{equation}    
The densities  $n_0$ and $n_1$ obtained so far are  for just one 
 D2 brane.  
For $N$ D2 branes  
with the same noncommutativity 
on each brane, which will be constructed below,  
a similar computation leads  
to 
\begin{equation}     
n_0={N\over 2\pi \theta}\,,\ \ \  
n_1= {N\theta\over (2\pi)^2 g_s l^3_s} 
\label{d0}     
\end{equation}   
Thus we obtain relations 
\begin{equation}     
N^2= n_0 n_1 (2\pi)^3 g_s l^3_s\,, 
\ \ \  \theta^2 = 
{n_1\over n_0} 2\pi g_s l_s^3\,. 
\label{mbounds}     
\end{equation}   
These are for the case where 
 all the strings and D0's are used up constructing 
the D2 branes without any extra D0's or strings. 
The first equation in (\ref{mbounds}) agrees with the saturated 
supergravity bound (\ref{bound}). The second equation can be 
reproduced by looking at $B_{(2)}$ in the supergravity solution. 
One finds that $B^{(2)}_{xz} = (2 \pi l_s^2) B = \frac{k}{U}$. 
Using $\theta = \frac{1}{B}$ and  
looking at the near horizon region of  
$r \ll l_s (g_s n_1 R)^{1\over 5}$ for large $g_s R n_1 $, the second  
relation of (\ref{d0}) follows.   
Combining this with the 
relation between $N$ and $n_1$, $n_0$ in the saturated case, 
one gets perfect agreement 
with the second equation in (\ref{mbounds}). 
 
For the noncommutative Yang-Mills theory description of the  
worldvolume dynamics,  
we introduce  gauge fields by 
\begin{equation} 
X=x+\theta A_z\,,\ \ \ 
Z=z-\theta A_x\,. 
\end{equation} 
Using $[x, \ ]=i\theta\partial_z$ and   
$[z, \ ]=-i\theta\partial_x$, one gets   
\begin{equation}     
[X, Z]= i\theta^2 \left({1\over \theta}+ F_{x z}\right)\,.    
\label{magnetic}     
\end{equation}  
Thus it is clear that the worldvolume background magnetic   
field is  
\begin{equation}     
B={1\over \theta}\,.    
\label{backmagnetic}     
\end{equation} 
In order to evaluate the background electric field,    
we shall use the relation $D_0 X= -\theta E_z$ 
and $D_0 Z= \theta E_x$.  
Evaluated on the explicit solution, one finds that  
\begin{equation}     
E_x=0,\ \ \  E_z= -{1\over 2\pi \alpha'} 
\,.    
\label{bakelectric}     
\end{equation}  
From this, we conclude that the  background electric 
field in the z-direction is ``critical''.

The analysis of worldvolume gauge theory 
is straightforward and simpler than that of the supersymmetric tubes.  
To organize the worldvolume theory, one may use  
 either a standard noncommutative gauge theory  with the  
background electric field 
or a deformed noncommutative gauge theory   
without any background. 
Here we shall take the latter approach, which turns  
out to be 
more convenient for a D2 brane with electric flux. 
First we define the worldvolume gauge field ${\cal A}_\mu$ by 
${\cal A}_t=A_0-Z$ and 
$X=x+\theta {\cal A}_z$, $Z=z-\theta {\cal A}_x$ 
and $Y=\theta\, \varphi$. 
Inserting this into the original  
matrix model 
and ignoring total derivative terms,  
one is led to 
\begin{equation}     
L={\theta^2\over 2}\tr \left( {\cal F}^2_{tx}+ 
({\cal F}_{tz}-\theta {\cal F}_{xz})^2  
-\theta^2 {\cal F}^2_{xz} 
+(\nabla_t\varphi-\theta\nabla_x\varphi)^2-\theta^2\Bigl( 
(\nabla_x\varphi)^2+(\nabla_z\varphi)^2\Bigr) 
\right)\,,    
\label{world1}     
\end{equation} 
where 
\begin{equation}    
\nabla_\mu=\partial_\mu-i[{\cal A}_\mu,\ ]\,,\ \ \  
{\cal F}_{\mu\nu}=\partial_\mu {\cal A}_\nu- 
\partial_\nu {\cal A}_\mu-i[{\cal A}_\mu,{\cal A}_\nu]\,.  
\end{equation}    
In this worldvolume action, the metric takes a rather unusual form, 
which has  nonvanishing off diagonal elements.  
However as we will verify  in the next section,  
the metric appearing in the  
action is the natural open string metric induced from  
the matrix theory description. 
 
By redefining  $x\rightarrow  
x +\theta t$ and ${\cal A}_t \rightarrow {\cal A}_t- 
\theta {\cal A}_x$, the Lagrangian becomes 
\begin{equation}     
L={\theta^2\over 2}\tr \left( ({\cal F}_{tx})^2+({\cal F}_{tz})^2 
-\theta^2 {\cal F}^2_{xz} 
+(\nabla_t\varphi)^2-\theta^2\Bigl( 
(\nabla_x\varphi)^2+(\nabla_x\varphi)^2\Bigr) 
\right)\,.    
\label{lagstatic}     
\end{equation} 
One could add also the contributions of the remaining 6 transverse  
scalars. 
Rescaling the time coordinate  
$t\rightarrow {\theta\over 2\pi\alpha'} t$, one can make the  
worldvolume theory 
to be the noncommutative Yang-Mills  
theory with a standard flat metric $\eta_{\mu\nu}$. 
Or one may rescale the spatial coordinate to get the 
noncommutative Yang-Mills  
theory with a standard flat metric $\eta_{\mu\nu}$. 
 
 
We now move to the case of ${\rm D2}$-$\overline{\rm D2}$. 
The solution is 
\begin{eqnarray} 
X+iZ&=&\sqrt{2\theta}\sum^\infty_{n=0}\sqrt{n+1} \Bigl(|2n\ket\bra 2n+2|+ 
|2n+3\ket\bra 2n+1|\Bigr) 
\,,\nonumber\\ 
 Y&=&{\xi\over 2}\sum^\infty_{n=0}\Bigl( 
|2n\ket\bra 2n|-|2n+1\ket\bra 2n+1|\Bigr)\,. 
\end{eqnarray} 
The basis labelled by the even numbers describes a D2 
brane  located at $y=\xi/2$. This brane is extended in the $xz$ directions 
and its worldvolume background fields 
are $B=1/\theta$ and $E_z=-1/(2\pi\alpha')$. 
On the other hand, the odd-number basis is for an anti-D2 brane 
extended again $xz$ directions but located at $y=-\xi/2$. 
The worldvolume background fields of the anti-D2 brane 
are $B=-1/\theta$ and $E_z=-1/(2\pi\alpha')$. 
Thus the branes are separated by $\xi$ in the $y$ direction. 
This is the configuration of ${\rm D2}$-$\overline{\rm D2}$ 
system obtained by taking the degenerate planar limit of the  
supersymmetric tube.

To determine the related charges, we use (\ref{charges}). 
From the first term, the number of D0's are  
again given by $\tr I$. The third term vanishes  
and the net D2 charge is zero. 
The moment is now nonvanishing and evaluated by 
\begin{equation}     
S^{\rm dipole}_{\rm CS}=  
{i\mu_0\lambda^2\over 3} \int\, dt\, \tr  
\phi^I\phi^J\phi^K F^{(4)}_{t\,IJK}=  
-{1\over 3}{\xi\over(2\pi)^2  g_s \,\,l_s^3}   
\int dt dx dz    
F^{(4)}_{txzy}  
\,.   
\end{equation} 
Here we have used $2\pi \theta \tr' =\int dxdz$ 
where $\tr'$ is  the trace operation over the matrices 
with basis $|n\ket'\bra m|'$.        
The dipole moment density  is then  
\begin{equation}     
d_2= {1\over 3}  
{\xi\over (2\pi)^2 g_s\,\, l_s^3}={1\over 3}\,\mu_2\, \xi\,.   
\end{equation}      
The dipole moment is proportional to the transverse separation of 
the ${\rm D2}$ and $\overline{\rm D2}$ and agrees with that of 
the supertube\cite{swkim}.

For the worldvolume description, the language 
of U(2) gauge theory is more appropriate.  
The $U(2)$ basis can be constructed by writing  
$|2n-1+a\ket \bra 2m-1 +b|= 
|n\ket'\bra m|' \tau_{ab}$ ($a,b=1,2$).    
Here $|n\ket'$ is  
interpreted as a new basis for the  
space while $\tau_{ab}$ generates an 
$U(2)$ algebra.  
We further introduce the notation 
$C\equiv {1\over \sqrt{2}} (X +i Z)$ 
with  
\begin{eqnarray} 
{C}= 
\left( 
\begin{array}{cc} 
\!\!U & W\\ 
T^\dagger & V^\dagger  
\end{array} 
\right) = \tau_{11}\, U +\tau_{12}\, W + \tau_{21}\, T^\dagger +  
\tau_{22}\, V^\dagger  
\end{eqnarray} 
where  
U, V, T and  W are $\infty\times\infty$ matrices 
of basis $|n\ket'\bra m|'$.  
The ${\rm D2}$-$\overline{\rm D2}$ background is  
 described by 
\begin{eqnarray} 
\bar{C}= 
\left( 
\begin{array}{cc} 
\sqrt{\theta}\, c & 0\\ 
0 & \sqrt{\theta} c^\dagger  
\end{array} 
\right) = \tau_{11}\,\, \sqrt{\theta}\,c + \tau_{22}\,\, \sqrt{\theta}\, 
c^\dagger\,, 
\end{eqnarray} 
and the U(2)-valued background field strength reads 
\begin{eqnarray} 
[\bar{C},\bar{C}^\dagger]=  
\theta (\tau_{11} - \tau_{22}).  
\end{eqnarray} 
As we are dealing with brane-antibrane system, 
the background magnetic field on each brane has opposite 
signature. 
For the worldvolume description, we organize the fluctuation  
as follows; 
\begin{eqnarray} 
{C}=\sqrt{\theta} 
\left( 
\begin{array}{cc} 
c & 0\\ 
0 & c  
\end{array} 
\right)+\sqrt{\theta} 
\left( 
\begin{array}{cc} 
0 & 0\\ 
0 & c^\dagger-c  
\end{array} 
\right) 
-{i\theta\over \sqrt{2}}({\cal A}_x+i{\cal A}_z) 
\end{eqnarray} 
The first term will generate the spatial derivatives 
and the second term is for the background gauge  
field,  
\begin{eqnarray} 
{\cal A}_x^{\rm back}= {2z\over \theta}\tau_{22},
\ \ \ {\cal A}^{\rm back}_z=0  
\end{eqnarray} 
which is nonvanishing only for the anti-D2 brane. 
The presence of such nontrivial  
background magnetic  
field is due to the fact that we describe  
anti D2-brane from the view point of D2-brane.  
In the ordinary ${\rm D2}$-$\overline{\rm D2}$,  
this background magnetic field  
makes the strings connecting    
${\rm D2}$ to $\overline{\rm D2}$ be tachyonic. To see this  
explicitly, 
let us turn on the $T$ and $W$ 
that describes ${\rm D2}$-$\overline{\rm D2}$  
strings and compute the potential 
\begin{eqnarray} 
\tr [{C},{C}^\dagger]^2=  
\tr'\left( (\theta \!+\!WW^\dagger\!-\!TT^\dagger)^2 
+(\theta \!+\!W^\dagger W\!-\!T^\dagger T)^2 
+2\theta |cT\!-\!Tc^\dagger\! +\!W c\!-\!c^\dagger W|^2 
\right)\,.  
\end{eqnarray} 
As is well known, 
certain components of $T$ have a negative mass squared and become  
tachyonic in case of the ordinary ${\rm D2}$-$\overline{\rm D2}$  
system\cite{shenker,li,mandal}. 
For example, the mode $T=u |0\ket'\bra 0|'$ has a quadratic  
potential term $-4 \theta u^2$, so it is tachyonic. 
In the ordinary ${\rm D2}$-$\overline{\rm D2}$  system,  
the vortex-antivortex  annihilation 
is argued to be an important process 
for the  decay of the tachyons\cite{shenker,li,mandal}. 
In our case, the  
vortices (D0's)  
are also stable  
objects as we will see later on. 
 
Now we like to prove that there are no tachyons in our ${\rm D2}$-$\overline{\rm D2}$  
system. 
For this, we proceed as follows. 
First using ${\cal A}_t=A_0-Z$,  
rewrite the original matrix  Lagrangian by 
\begin{equation}     
L={1\over 2 } \tr \left( (\nabla_0 X)^2  + 
(\nabla_0 Y)^2 +(\nabla_0 Z)^2        
+[X,Y]^2 +2i \nabla_0 X [X,Z]  
+2i \nabla_0 Y [Y,Z]  
\right)    
\label{lag2}     
\end{equation}  
where $\nabla_0=\partial_0-i[{\cal A}_0,\ ]$ 
and we suppress the contributions of 
the remaining scalars. 
Now we compute the corresponding Hamiltonian by 
the Legendre transform; the resulting expression  
reads 
\begin{equation}     
\tilde{E}={1\over 2 } \tr \left( (\nabla_0 X)^2  + 
(\nabla_0 Y)^2 +(\nabla_0 Z)^2        
+|[X,Y]|^2  
\right)    
\label{hamiltonian2}     
\end{equation}  
which is in fact $H-\tr C_J$ with $\tr C_J$ 
being the central charge in (\ref{centralcharge}). 
This Hamiltonian is dynamically equivalent to the original Hamiltonian 
and equally well describes the  
dynamics of the matrix model. 
We now evaluate contributions of  
any fluctuations around the BPS background using this new  
Hamiltonian.  
It is obvious to see that 
the leading contributions are quadratic in 
 fluctuations and positive definite. 
The energy in total is positive definite, so  
 any fluctuations cost energy. In conclusion there are no 
tachyons, 
as it should be since we are considering fluctuations 
around the BPS background.

The more general BPS solutions for the collection of parallel  
${\rm D2}$'s and  
$\overline{\rm D2}$'s are 
\begin{eqnarray} 
&&X={1\over \sqrt{2}}\sum^{N-1}_{a=0}u_a\sum^\infty_{n=0}\sqrt{n+1}  
\Bigl(|N(n+1)+a\ket\bra Nn+a| 
+|Nn+a\ket\bra N(n+1)+a|\Bigr) 
\,,\nonumber\\ 
&&Z={1\over \sqrt{2}}\sum^{N-1}_{a=0}v_a 
\sum^\infty_{n=0}i\sqrt{n+1}  
\Bigl( 
|N(n+1)+a\ket\bra Nn+a| 
-|Nn+a\ket\bra N(n+1)+a| 
\Bigr)\,,\nonumber\\ 
&& Y=\sum^{N-1}_{a=0}y_a 
\sum^\infty_{n=0} 
|Nn+a\ket\bra Nn+a|\,, 
\end{eqnarray}  
where the integer $N$ is the total  number of ${\rm D2}$ and  
$\overline{\rm D2}$  
and $u_a$, $v_a$ and $y_a$ are all real parameters. 
The parameter $y_a$ represents  
the transverse location of the $a$-th D2-brane. 
$|u_a v_a|$ (no sum) is the noncommutative parameters 
on $a$-th brane while the signature of  $u_a v_a$ 
indicates whether the $a$-th brane is ${\rm D2}$  or    
$\overline{\rm D2}$. From our convention, the positive  
signature represents D2.

Here is an example of more  
nontrivial configuration;  
\begin{eqnarray} 
&&X={1\over \sqrt{2}}\sum^\infty_{n=0} 
\Bigl[ u_0\sqrt{n+1} \left(|2n+2\ket\bra 2n| 
+|2n\ket\bra 2n+2|\right)\Bigr.\nonumber\\ 
&&\ \ \ \ \ \ \ \ \ \ \ \ \ \ \ \   
+ u_1 \cos\chi\,\sqrt{n+1} \left(|2n+3\ket\bra 2n + 1| 
+|2n+1\ket\bra 2n+3|\right) 
\Bigr] 
\,,\nonumber\\ 
&&  
Y={u_1\,\sin\chi\, \over \sqrt{2}}\sum^\infty_{n=0}\sqrt{n+1}  
\Bigl(|2n+3\ket\bra 2n+1| 
+|2n+1\ket\bra 2n+3|\Bigr)\,,\nonumber\\ 
&&Z={1\over \sqrt{2}}\sum^{1}_{a=0}v_a 
\sum^\infty_{n=0}i\sqrt{n+1}  
\Bigl( 
|2(n+1)+a\ket\bra 2n+a| 
-|2n+a\ket\bra 2(n+1)+a| 
\Bigr) 
\,. 
\end{eqnarray}  
This solution  describes a configuration in which  
the 0th and 1st branes are  extended respectively  
in the directions 
 $txz$ and $tx'z$ with $x'= \cos\chi\, x + \sin\chi\, y$.  
Namely 
the two D2-branes make an arbitrary angle $\chi$ 
in the $xy$ plane 
 with $z$  
as a common direction. 
The solution is again 1/4 BPS but 
(\ref{bps}) is not an appropriate  BPS equation 
for them.  Rather they satisfy the original BPS equations  
in (\ref{bpseq}).  

There is yet another intriguing class of BPS solutions for  
Eq. (\ref{bps}); an example is  
\begin{eqnarray} 
&& X={\rho\over 2}\sum^\infty_{n=-\infty}(|n+1\ket\bra n| 
+|n\ket\bra n+1| ) 
\,,\nonumber\\ 
&& Z=l\sum^\infty_{n=-\infty} n|n\ket\bra n|\,, 
\end{eqnarray} 
with $Y=0$. 
This solution is obtained from the supersymmetric  
tube solution\cite{lee} 
by setting $Y=0$. There are no net D2-brane charges 
nor  the dipole moment carried by this object. 
How to interpret this solution seems not clear. 
Perhaps  a simple minded interpretation will be 
 an elliptic tube 
with the limit where the length of  minor axis  
becomes zero. 
Indeed when we diagonalize $X$, the magnitude of its eigenvalues 
are bounded by $\rho$. This  indicates that we are dealing with 
a strip having a width $\rho$ in the x-direction,  
extended infinitely to the z-direction.    
The detailed worldvolume description of this kind of object is not  
 known 
though the fluctuation analysis of the matrix model 
can in principle provide it.

\section{Open string metric} 
 
Here we like to compute the metric for the noncommutative worldvolume  
theory on the D2. For this, we shall follow the procedure described  
in  
Ref.\cite{seiberg}. 
We begin with the closed string metric of the diagonal form 
$g_{\mu\nu}= {\rm diag}(-|g_{tt}|, g_{xx}, g_{zz})$ and  
$B_{\mu\nu}$ of the form, 
\begin{eqnarray} 
B_{\mu\nu}=\left( 
\begin{array}{ccc} 
0 & 0 & E \\ 
0 & 0 & B \\ 
- E & -B & 0  
\end{array} 
\right) 
\end{eqnarray} 
which is nothing but the $B_{\mu\nu}$ for the D2-brane. 
With the metric and after restoring $2\pi\alpha'$, 
the conditions for supersymmetry, (\ref{solveone}) and 
(\ref{dzero}) together with (\ref{critical}) read 
\begin{eqnarray} 
|g_{tt}| g_{xx}g_{zz}= g_{xx} (2\pi\alpha' E)^2\,, 
\label{electric} 
\end{eqnarray} 
and, hence, $\lambda E=\pm \sqrt{|g_{tt}|g_{zz}}$. 
The open string metric and $\theta$ can be identified using the  
following relation\cite{seiberg}: 
\begin{eqnarray} 
{1\over g+\lambda B}= {\theta\over\lambda} + {1\over G+\lambda \Phi}, 
\label{seiberg}  
\end{eqnarray} 
where  $G$ and $\theta^{\mu\nu}$ are respectively the open string metric and 
the noncommutativity of the worldvolume theory. The two form  
$\Phi$ is free to  
choose but there is natural one for the matrix theory  
description\cite{seiberg}. 
With the value $E$ in (\ref{electric}), note 
\begin{eqnarray} 
{1\over g+\lambda B} 
=-{1\over |g_{tt}|(\lambda B)^2 }\left( 
\begin{array}{ccc} 
g_{xx}g_{zz}+(\lambda B)^2 &  -\lambda B \lambda E & -g_{xx} \lambda E \\ 
-\lambda B \lambda E & 0 & |g_{tt}|\lambda B \\ 
 g_{xx} \lambda E & -|g_{tt}|\lambda B  & -|g_{tt}|g_{xx}  
\end{array} 
\right) 
\end{eqnarray} 
The Seiberg-Witten limit\cite{seibergwitten} corresponds to 
\begin{eqnarray} 
\lambda \sim \sqrt{\epsilon}\,,\ \ \ 
g_{xx}\sim g_{zz}\sim \epsilon 
\end{eqnarray} 
keeping $B$, $E$ and $g_{tt}$ fixed. 
In this limit, one finds that 
\begin{eqnarray} 
\theta=\left( 
\begin{array}{ccc} 
0 & 0 & 0 \\ 
0 & 0 & -1/B \\ 
0 & 1/B & 0  
\end{array} 
\right)\,, 
\end{eqnarray} 
for any fixed $\Phi$. 
One can go forward to compute the corresponding $G_{\mu\nu}$ or the inverse  
$G^{\mu\nu}$; the result agrees with the expression  
below. 
 
Alternatively, from the matrix theory description we know that 
\begin{eqnarray} 
\theta=-\Phi^{-1}=\left( 
\begin{array}{ccc} 
0 & 0 & 0 \\ 
0 & 0 & -1/B \\ 
0 & 1/B & 0  
\end{array} 
\right)\,. 
\end{eqnarray} 
Those are the noncommutativity and $\Phi$ used for  
the worldvolume theory. From this, one may compute $G_{\mu\nu}$  
without taking the Seiberg-Witten limit. From (\ref{seiberg}), they are 
\begin{eqnarray} 
G_{\mu\nu}=\left((g+\lambda B)^{-1}- {\theta\over\lambda}\right)^{-1}   
-\lambda \Phi\,. 
\end{eqnarray} 
The straightforward evaluation gives 
\begin{eqnarray} 
G_{\mu\nu}= 
\left( 
\begin{array}{ccc} 
0 & \pm \lambda  B \sqrt{|g_{tt}|\over g_{zz}} & 0 \\ 
\pm \lambda B \sqrt{|g_{tt}|\over g_{zz}} & {(\lambda B)^2\over  
g_{zz}} & 0 \\ 
0 & 0 &  {(\lambda B)^2\over g_{xx}} 
\end{array} 
\right)\,, 
\end{eqnarray} 
and  
\begin{eqnarray} 
G^{\mu\nu}= 
\left( 
\begin{array}{ccc} 
-{1\over |g_{tt}|} & \pm  {\sqrt{g_{zz}}\over \lambda B \sqrt{|g_{tt}|}} & 0 \\ 
\pm  {\sqrt{g_{zz}}\over \lambda B \sqrt{|g_{tt}|}} & 0 & 0 \\ 
0 & 0 &  {g_{xx}\over (\lambda B)^2} 
\end{array} 
\right)\,. 
\end{eqnarray} 
This $G^{\mu\nu}$ is precisely the  
metric appearing 
in the action  
(\ref{world1}). 
Here we do not have to take  the Seiberg-Witten limit,  
but the same metric  
follows from the limit as mentioned before. 
 

\section{D0-D2 solutions} 
The supergravity solutions implies that one could have extra  
D0's which are not used up  to form D2 branes. The configurations 
are again 1/4 BPS. In this section, we shall briefly discuss 
such solutions. Since generalization to the case of 
N D2-brane is straightforward, we shall restrict our discussion 
to the case of one D2 brane. 
As done for the case supersymmetric tubes, 
we  introduce a shift operator defined by  
\begin{eqnarray}     
S = \sum^\infty_{n=0} |n+m\ket \bra n|\,.   
\end{eqnarray}       
It satisfies the relations   
\begin{eqnarray}     
S S^\dagger= I-P\,,\ \ \    
S^\dagger S =I\,,   
\end{eqnarray}        
where the projection operator $P$ is defined by $  
P=\sum^{m-1}_{a=0} |a\ket \bra a| $.   
Then general soliton solutions including the moduli parameters  
are given by \cite{bak, park}  
\begin{equation}   
 {X}_i=S \, x_i S^\dagger +  
\sum^{m-1}_{a=0}\xi_i^{a}|a\ket\bra a| \, , \,\,\,\,   
 {X}_s= \sum^{m-1}_{a=0}\varphi_s^{a}|a\ket\bra a|\,,   
\label{tubed0}     
\end{equation}  
where $i=1,2$ are respectively for $x$ and $z$, i.e.
$x=x_1,\, z= x_2$  
and  the index $s$ refers to the remaining seven  
transverse scalars. 
This certainly satisfies the  
BPS equations.  
Hence the solution is again 1/4 BPS and the presence 
of solitons does not break any further supersymmetries. 
The solutions describes $m$ extra D0 branes 
with there positions ($\xi^a_i$, $\varphi^a_s$) in the nine  
dimensional target space.  
Using the field defined in (\ref{lagstatic}), the solution becomes 
\begin{eqnarray}   
&& {\cal A}_i= -{1\over \theta}\epsilon_{ij}\left(S \, x_j S^\dagger -x_j + 
\sum^{m-1}_{a=0}\left(\xi_j^{a}+{\theta t\over 2\pi\alpha' }\delta_{j1} 
\right)|a\ket\bra a| \right)\nonumber\\   
&& {\varphi}_s={1\over \theta}  
\sum^{m-1}_{a=0}\varphi_s^{a}|a\ket\bra a|\,,   
\label{tubed000}     
\end{eqnarray}  
in an appropriate gauge. 
The field contents of the solution  are identified as  
\begin{equation}   
{\cal F}_{tz}={1\over 2\pi \alpha'}P\,, \ \ \   
{\cal F}_{xz}=-{1\over \theta}P\,. 
\label{fields}     
\end{equation}  
 
This may be compared with the moving solitons found in \cite{park}. 
It is then clear that this corresponds  
to moving D0's with velocity  
$v_x={\theta\over 2\pi\alpha'}$.  
In fact the bosonic content of the Lagrangian in (\ref{lagstatic}) 
is exactly the same as the noncommutative Yang-Mills theory describing 
D2 brane worldvolume in the decoupling  
limit. 
We know that the D0's on D2 with the  
noncommutativity turned on are   
tachyonic, which was explicitly verified in \cite{aganagic,ohta}. 
 Hence we are led to a seemingly contradictory  
result since the above solitons should be stable  
as the remaining 
supersymmetries dictate. 
However there is one way to avoid this conclusion; since in our case 
the D0's are moving in a specific velocity, this specific motion 
can make the tachyonic spectrum disappear.  
Indeed, one can prove that $\tr {\cal F}_{tz}$ is conserved in  
general  using the equations of motion. Hence it is not possible 
to dynamically reduce the velocity to make them unstable, which confirms 
that there is no contradiction.  
 

One may ask what happens to the worldvolume 
 solitons for the 1/4 BPS N Dp  or  Dp anti-Dp systems 
that can be obtained by the T-dualization along 
the transverse directions.  Our analysis above can be trivially 
extended to the case of noncommutative solitons describing 
branes of codimension two. It is, however, not so straightforward  
if one considers  
worldvolume solitons  
of D$p'$ with $p'<p-2$.  
 For example, how magnetic monopoles corresponding to D-strings
connecting D3-branes get affected due to the change of 
the worldvolume theory, seems quite interesting.
Or one could study the possible deformation of instantons. 
Though we believe these are  interesting issues, we like to leave them 
for the future work, partly due to the vastness of the 
subject itself.

\section{Conclusion} 
 
In this note, we have obtained the supersymmetric   
brane-antibrane configurations. 
The configurations may be obtained by taking 
a degenerate limit of the elliptic deformation of  
the supersymmetric  
tubes. We have verified that 
 the systems preserve 8 supersymmetries via the analysis 
of the worldvolume kappa-symmetry or the matrix theory 
and constructed corresponding supergravity solutions. 
The specific  worldvolume background  
electric field, which is induced by the  
 dissolved fundamental strings on the branes, 
makes the would-be tachyonic degrees  
disappear. We have shown that the worldvolume dynamics 
may be described by  gauge theories with the  
spatial noncommutativity.   
Finally we  
study D0-solitons on the D2 brane, 
which is supersymmetric and stable. 
 
The worldvolume background gauge fields   
involve the magnetic as well as electric components. 
From this one may naively expect that the natural  
description would  
be spacetime noncommutative. However as we have constructed  
in detail, 
the study of the matrix model leads to the worldvolume  
gauge theory with 
only spatial noncommutativity. The metric appearing in the  
gauge theory is shown to agree precisely  
with the expected open string metric of the  
string theory. Note, however, that 
the only invariant combination  
$\lambda^2 F^2= g^{tt}g^{zz}(\lambda  
E_z)^2 + 
g^{xx}g^{zz} (\lambda B)^2$ may flip the signature depending on  
the magnitude of $\theta=1/B$. With the specific value of $E_z$  
in (\ref{electric})  
required by the supersymmetry, the invariant combination 
becomes $\lambda^2 F^2= -1+ 
g^{xx}g^{zz} (\lambda/\theta)^2$.  
Depending on the value of the noncommutativity scale  
$\theta$, $F^2$ may be spacelike, lightlike or  
timelike. Yet in the worldvolume theory,  
there appears no apparent signals indicating  
possible breakdown\cite{gomis} or  
transitions of  
the theories.   Is the worldvolume noncommutative  
theory we obtained somehow related to the lightlike noncommutative  
theory in Refs.\cite{gomis,aharony} when $F^2=0$?  
Currently, the issues here are not fully resolved. 
One thing clear is that  
our description of the worldvolume theory 
can be trusted if one takes 
the Seiberg-Witten decoupling limit\cite{seibergwitten}. 
In this limit, $F^2$ is dominated by the $B^2$ term 
and thus spacelike always. 
 
Finally, we like to comment upon the nature 
of the worldvolume theory when we have many parallel 
D2-branes with varying noncommutative  
scales.  
This is contrasted to the case of 1/2 BPS branes, where the BPS condition  
dictates all the noncommutativity scale the same\footnote{Nonetheless, 
one could think of varying noncommutativity scale in the transverse  
directions which leads to the mixing of the nonabelian 
symmetry and geometry\cite{yin}.}. 
The original matrix model has a $U(N)$ noncommutative  
gauge symmetry, $X_I\rightarrow U^\dagger  X_I U $, if one considers  
N such branes. Since the noncommutativity scale controls the natural 
open string metric and the Yang-Mills coupling constant on each brane, 
this gauge symmetry is broken by the presence of such branes 
with varying noncommutativity scales.  
Of course one could separate the branes in the transverse  
directions, which leads to the conventional way of spontaneous 
breaking of the nonabelian gauge symmetry. 
Being independent of the transverse separation, 
the breaking  induced by the varying noncommutativity scale  
makes  the open string geometries and the couplings vary 
from one to another branes. 
Further investigation is necessary on the interplay between 
the geometries and the nonabelian symmetry.

\noindent{\large\bf Acknowledgment}  
DB would like to thank Kimyeong Lee and Soo-Jong Rey for 
discussions. AK would like to thank Michael Gutperle and Shiraz Minwalla.    
This work is supported in part by KOSEF 1998    
Interdisciplinary Research Grant 98-07-02-07-01-5 
and by NSF grant PHY 9218167.


\bibliography{bak} 

\begingroup\raggedright\begin{thebibliography}{10}

\bibitem{mateos}
D.~Mateos and P.~K. Townsend, ``Supertubes,'' {\em Phys. Rev. Lett.} {\bf 87}
  (2001) 011602, \href{http://xxx.lanl.gov/abs/hep-th/0103030}{{\tt
  hep-th/0103030}}.

\bibitem{lee}
D.~Bak and K.~Lee, ``Noncommutative supersymmetric tubes,'' {\em Phys. Lett.}
  {\bf B509} (2001) 168--174,
  \href{http://xxx.lanl.gov/abs/hep-th/0103148}{{\tt hep-th/0103148}}.

\bibitem{emparan}
R.~Emparan, D.~Mateos, and P.~K. Townsend, ``Supergravity supertubes,'' {\em
  JHEP} {\bf 07} (2001) 011, \href{http://xxx.lanl.gov/abs/hep-th/0106012}{{\tt
  hep-th/0106012}}.

\bibitem{swkim}
D.~Bak and S.-W. Kim, ``Junctions of supersymmetric tubes,''
  \href{http://xxx.lanl.gov/abs/hep-th/0108207}{{\tt hep-th/0108207}}.

\bibitem{bergshoeff}
E.~Bergshoeff and P.~K. Townsend, ``Super D-branes,'' {\em Nucl. Phys.} {\bf
  B490} (1997) 145--162, \href{http://xxx.lanl.gov/abs/hep-th/9611173}{{\tt
  hep-th/9611173}}.

\bibitem{cho}
J.-H. Cho and P.~Oh, ``Super D-helix,''
  \href{http://xxx.lanl.gov/abs/hep-th/0105095}{{\tt hep-th/0105095}}.

\bibitem{lunin}
O.~Lunin and S.~D. Mathur, ``Metric of the multiply wound rotating string,''
  {\em Nucl. Phys.} {\bf B610} (2001) 49--76,
  \href{http://xxx.lanl.gov/abs/hep-th/0105136}{{\tt hep-th/0105136}}.

\bibitem{banks1}
T.~Banks, W.~Fischler, S.~H. Shenker, and L.~Susskind, ``M theory as a matrix
  model: A conjecture,'' {\em Phys. Rev.} {\bf D55} (1997) 5112--5128,
  \href{http://xxx.lanl.gov/abs/hep-th/9610043}{{\tt hep-th/9610043}}.

\bibitem{seiberg1}
N.~Seiberg, ``Why is the matrix model correct?,'' {\em Phys. Rev. Lett.} {\bf
  79} (1997) 3577--3580, \href{http://xxx.lanl.gov/abs/hep-th/9710009}{{\tt
  hep-th/9710009}}.

\bibitem{banks}
T.~Banks, N.~Seiberg, and S.~H. Shenker, ``Branes from matrices,'' {\em Nucl.
  Phys.} {\bf B490} (1997) 91--106,
  \href{http://xxx.lanl.gov/abs/hep-th/9612157}{{\tt hep-th/9612157}}.

\bibitem{myers}
R.~C. Myers, ``Dielectric-branes,'' {\em JHEP} {\bf 12} (1999) 022,
  \href{http://xxx.lanl.gov/abs/hep-th/9910053}{{\tt hep-th/9910053}}.

\bibitem{shenker}
P.~Kraus, A.~Rajaraman, and S.~H. Shenker, ``Tachyon condensation in
  noncommutative gauge theory,'' {\em Nucl. Phys.} {\bf B598} (2001) 169--188,
  \href{http://xxx.lanl.gov/abs/hep-th/0010016}{{\tt hep-th/0010016}}.

\bibitem{li}
M.~Li, ``Note on noncommutative tachyon in matrix models,'' {\em Nucl. Phys.}
  {\bf B602} (2001) 201--212,
  \href{http://xxx.lanl.gov/abs/hep-th/0010058}{{\tt hep-th/0010058}}.

\bibitem{mandal}
G.~Mandal and S.~R. Wadia, ``Matrix model, noncommutative gauge theory and the
  tachyon potential,'' {\em Nucl. Phys.} {\bf B599} (2001) 137--157,
  \href{http://xxx.lanl.gov/abs/hep-th/0011094}{{\tt hep-th/0011094}}.

\bibitem{seiberg}
N.~Seiberg, ``A note on background independence in noncommutative gauge
  theories, matrix model and tachyon condensation,'' {\em JHEP} {\bf 09} (2000)
  003, \href{http://xxx.lanl.gov/abs/hep-th/0008013}{{\tt hep-th/0008013}}.

\bibitem{seibergwitten}
N.~Seiberg and E.~Witten, ``String theory and noncommutative geometry,'' {\em
  JHEP} {\bf 09} (1999) 032, \href{http://xxx.lanl.gov/abs/hep-th/9908142}{{\tt
  hep-th/9908142}}.

\bibitem{bak}
D.~Bak, ``Exact multi-vortex solutions in noncommutative Abelian- Higgs
  theory,'' {\em Phys. Lett.} {\bf B495} (2000) 251--255,
  \href{http://xxx.lanl.gov/abs/hep-th/0008204}{{\tt hep-th/0008204}}.

\bibitem{park}
D.~Bak, K.~Lee, and J.-H. Park, ``Noncommutative vortex solitons,'' {\em Phys.
  Rev.} {\bf D63} (2001) 125010,
  \href{http://xxx.lanl.gov/abs/hep-th/0011099}{{\tt hep-th/0011099}}.

\bibitem{aganagic}
M.~Aganagic, R.~Gopakumar, S.~Minwalla, and A.~Strominger, ``Unstable solitons
  in noncommutative gauge theory,'' {\em JHEP} {\bf 04} (2001) 001,
  \href{http://xxx.lanl.gov/abs/hep-th/0009142}{{\tt hep-th/0009142}}.

\bibitem{ohta}
A.~Fujii, Y.~Imaizumi, and N.~Ohta, ``Supersymmetry, spectrum and fate of D0-Dp
  systems with B- field,'' \href{http://xxx.lanl.gov/abs/hep-th/0105079}{{\tt
  hep-th/0105079}}.

\bibitem{gomis}
J.~Gomis and T.~Mehen, ``Space-time noncommutative field theories and
  unitarity,'' {\em Nucl. Phys.} {\bf B591} (2000) 265--276,
  \href{http://xxx.lanl.gov/abs/hep-th/0005129}{{\tt hep-th/0005129}}.

\bibitem{aharony}
O.~Aharony, J.~Gomis, and T.~Mehen, ``On theories with light-like
  noncommutativity,'' {\em JHEP} {\bf 09} (2000) 023,
  \href{http://xxx.lanl.gov/abs/hep-th/0006236}{{\tt hep-th/0006236}}.

\bibitem{yin}
K.~Dasgupta and Z.~Yin, ``Non-Abelian geometry,''
  \href{http://xxx.lanl.gov/abs/hep-th/0011034}{{\tt hep-th/0011034}}.

\end{thebibliography}\endgroup
\bibliographystyle{ssg} 
\end{document}